\documentclass[aps,prd,twocolumn,superscriptaddress,preprintnumbers,floatfix,nofootinbib,notitlepage,showkeys,showpacs]{revtex4-1}

\usepackage[utf8]{inputenc}

\usepackage{graphicx}
\usepackage{hyperref}
\usepackage{latexsym}
\usepackage{amsmath}
\usepackage{amssymb}
\usepackage{bbm}
\usepackage{pdfsync}
\usepackage{epsfig}
\usepackage{epstopdf}
\usepackage{subfigure}
\usepackage{color}
\usepackage{comment}
\usepackage{slashed}
\usepackage{multirow}

\def\lsim{\mathrel{\raise.3ex\hbox{$<$\kern-.75em\lower1ex\hbox{$\sim$}}}}
\def\gsim{\mathrel{\raise.3ex\hbox{$>$\kern-.75em\lower1ex\hbox{$\sim$}}}}
\newcommand\CO{\text{CO}}
\newcommand\HI{\text{HI}}

\begin{document}

\title{Antisymmetric cross-correlation of line-intensity maps as a probe of reionization}

\author{Gabriela Sato-Polito}
\email{gsatopo1@jhu.edu}
\affiliation{Department of Physics and Astronomy, Johns Hopkins University, \\
Baltimore, MD 21218, USA}

\author{José Luis Bernal}
\affiliation{Department of Physics and Astronomy, Johns Hopkins University, \\
Baltimore, MD 21218, USA}

\author{Ely D. Kovetz}
\affiliation{Department of Physics, Ben-Gurion University of the Negev, Be’er Sheva 84105, Israel}

\author{Marc Kamionkowski}
\affiliation{Department of Physics and Astronomy, Johns Hopkins University, \\
Baltimore, MD 21218, USA}

\begin{abstract}
We present a new estimator for the cross-correlation signal between line intensity maps to probe the Epoch of Reionization. The proposed estimator is the hitherto neglected antisymmetric component of the cross-correlation, under the exchange of line-of-sight positions. We consider the cross-correlation between HI and CO fluctuations, and forecast the improvement in precision on reionization parameters when the antisymmetric contribution is accounted for. As a way to break the degeneracy between astrophysics and cosmology in the intensity mapping power spectrum, we study the ratio between the antisymmetric and symmetric components. While our results depend on the highly uncertain astrophysical modelling, we show that in most standard scenarios including the antisymmetric contribution as a complementary probe can lead to a significant gain in information. 

\end{abstract}

\maketitle

\section{Introduction}

Line-intensity mapping (LIM) is a technique that measures the integrated emission from atomic or molecular transitions of all sources along the line of sight~\cite{Kovetz:2017agg}. This can be used to measure the spatial fluctuations in the intensity of a given spectral line, with the radial position of the source determined by the frequency of the redshifted line. While observations of the cosmic microwave background (CMB) and galaxy surveys have mapped the early and late Universe with great precision~\cite{Aghanim:2018eyx, Alam:2016hwk}, large cosmological volumes at intermediate redshifts remain uncharted. At high redshifts galaxies become too faint and sparse, such that individual detections are insufficient for statistically significant cosmological measurements. 

LIM of different spectral lines across a wide range of redshifts will bridge between the volumes probed by CMB experiments and galaxy surveys (see e.g., \cite{Karkare:2018sar,2019PhRvL.123m1301M,Kovetz:2019uss,Silva:2019hsh,Bernal:2019gfq,Bernal:2020vbb}).
The 21-cm spin-flip transition in neutral hydrogen has been extensively studied as a probe of large-scale structure over a wide range of redshifts (see e.g.,~\cite{10.1093/mnras/188.4.791, 1990MNRAS.247..510S, Madau:1996cs,Mao:2008ug,Chang:2010jp, vanHaarlem:2013dsa, Tingay:2012ps,2015aska.confE...1K,Liu:2015gaa,DeBoer:2016tnn}). Along with these efforts, several other lines have been proposed as candidates for intensity mapping, such as the rotational lines from carbon monoxide (CO), [CII], H$\alpha$, and H$\beta$, with  particular attention given to the complementarity between different tracers (see e.g.,~\cite{Lidz:2008ry,Lidz:2011dx, Serra:2016jzs, 2017ApJ...848...52H}).  

The Epoch of Reionization (EoR) is a landmark transition in the history of the Universe that can be uniquely probed with LIM on cosmological scales. During this period, the first galaxies and quasars ionized the surrounding neutral hydrogen gas.  Upcoming measurements of the EoR will provide key insights into both astrophysics and cosmology. The intensity and distribution of emission lines during the EoR trace the underlying matter distribution and are sensitive to the astrophysical processes that took place. The LIM signal therefore promises to be an excellent probe of large-scale structure at high-redshifts, and will help elucidate the properties of the first stars and galaxies. Using different lines in conjunction, such as 21-cm and CO, will enable the mapping of both the neutral gas in the IGM and the galaxy distribution over the same cosmological volumes~\cite{Kovetz:2019uss,Silva:2019hsh}. Cross-correlation between these lines holds great promise for probing the onset and evolution of reionization~\cite{Lidz:2011dx}.

In its most general form, the two-point correlation function depends on the orientation and position of the two points. While it is often assumed that the correlation function is invariant under the exchange in position ${\bf r} \rightarrow -{\bf r}$, this does not necessarily hold for cross-correlations between different tracers of the matter density field. This asymmetry in the galaxy cross-correlation under the exchange of galaxy pairs has been studied both in Fourier and configuration space. Several potential contributions to the antisymmetric part of the cross-correlation have been pointed out, which include gravitational redshift \cite{McDonald:2009ud, Croft:2013taa}, gravitational lensing \cite{Jalilvand:2019bhk}, Doppler shift, light-cone effect, redshift evolution and the Alcock-Paczynski effect \cite{Bonvin:2013ogt,Gaztanaga:2015jrs}, as well as biased halo clustering, local-type primordial non-Gaussianity, early-Universe vector fields, etc.~\cite{Dai:2015wla}.

When applied to LIM, the antisymmetric component of the power spectrum is sensitive to the difference in redshift evolution of the temperature and bias of the cross-correlated lines. We may therefore expect a significant antisymmetric signal during the EoR, as a transition in the intergalactic medium is rapidly progressing. On the other hand, line-intensity fluctuations carry information about astrophysics and cosmology, and disentangling them is one of the main challenges to LIM observations (see e.g., Ref.~\cite{Bernal:2019jdo}). Here we propose two estimators in order to address these challenges:  the LIM antisymmetric cross-correlation estimator and the ratio between the antisymmetric and symmetric components. We study the detectability of both estimators and forecast their sensitivity to reionization parameters as a proof of concept.

We present a general framework for the antisymmetric and symmetric components of the LIM angular power spectrum. Since the amplitude of the antisymmetric signal depends on the bias and temperature evolution with redshift, the only requisite for the choice of spectral lines is that they must evolve unevenly over redshift. We choose to study the fluctuations in the intensity of the emission produced by the spin-flip transition of neutral hydrogen (which we refer to as HI line), since they probe the spatial structure of reionization directly. We cross-correlate the intensity of this line with the CO(2-1) rotational transition, an excellent tracer of star formation.

Assuming simple analytical models for the line emissions, we forecast the uncertainties on reionization parameters for futuristic LIM experiments. We find that the antisymmetric and symmetric components of the cross-correlation have different degeneracies and therefore lead to complementary constraints on the EoR. Although the precise gain in information from considering the antisymmetric cross-correlation is highly model-dependent, we show that it can be an important additional probe in most scenarios, especially with low noise. 

This paper is organized as follows. In Section~\ref{sec:limPS} we describe our approach  to model the LIM power spectrum and its noise. In Section~\ref{sec:estimator} we present the novel antisymmetric estimator, as well as the estimator for the symmetric part, and the ratio between them. We then discuss the covariance for each estimator. The detectability for a particular astrophysical model and survey configuration are shown in Section~\ref{sec:Forecast}. The precision to which each estimator can constrain the parameters that describe reionization, as well as how these constraints depend on the astrophysical modelling and instrumental noise, are shown in Section~\ref{sec:Forecast} as well. We conclude in Section~\ref{sec:conclusions}.

We adopt the standard $\Lambda$CDM cosmology throughout, with the following parameters from Planck 2018 \cite{Aghanim:2018eyx}: $h=0.674$, $\Omega_m = 0.315$, $\Omega_b = 0.049$, $n_s = 0.965$, and $\sigma_8 = 0.8$.

\section{LIM angular power spectrum}\label{sec:limPS}
Our fundamental observable is the spatial fluctuation of the brightness temperature of a given spectral line, defined as $\delta T \equiv T - \left\langle T \right\rangle$. The brightness temperature fluctuations can be projected on the sky and expanded using spherical harmonics. The angular power spectrum is then defined as the expectation value of the square of the spherical-harmonic coefficients.

The angular power spectrum between two tracers $X$ and $Y$ of the matter density field at redshift shells $z_i$ and $z_j$, respectively, is given by
\begin{equation}
    C^{X,Y}_{\ell}(z_i, z_j) = 4\pi \int \frac{dk}{k} \Delta^{X, z_i}_{\ell}(k) \Delta^{Y, z_j}_{\ell}(k) \mathcal{P}(k),
\end{equation}
where $\mathcal{P}(k)$ is the dimensionless matter power spectrum today and $\Delta^{X, z_i}_{\ell}(k)$ is the observed transfer function.  We include in the definition of the observed transfer function for LIM fluctuations the smoothing due to the limited angular resolution of LIM experiments. This can be modeled as an effective Gaussian beam $B^X_{\ell}$ that smooths the temperature fluctuations on small scales, restricting the number of accessible modes. Similarly, the spectral resolution would smooth modes along the line-of-sight. This contribution can be neglected as long as the redshift bins are larger than the width of the frequency channel. 

The observed transfer function is therefore given by 
\begin{equation}
    \Delta^{X, z_i}_{\ell}(k) = \int dz B^X_{\ell} W^X(z,z_i) \Delta^X_{\ell}(k,z),
\end{equation}
where 
\begin{equation}
    B^X_{\ell} = \exp\left( -\frac{\ell (\ell+1) (\theta^X_{\rm FWHM})^2}{16 \log 2} \right),
\end{equation}
$\theta^X_{\rm FWHM}$ is the full width at half maximum of the beam profile of the experiment targeting the spectral line $X$, and $W^{X}(z, z_i)$ is a normalized window function centered on $z_i$ which we assume to be a Gaussian. The contribution  to the transfer function $\Delta^X_{\ell}(k,z)$ for a spectral line $X$ that is related with intrinsic clustering is given by

\begin{equation}
    \Delta^X_{\ell}(k,z) = \left\langle T^X \right\rangle(z) b^X(z) D(z) j_{\ell}[kr(z)],
    \label{eq:transfer}
\end{equation}
where $D(z)$ is the growth factor defined such that $D(0) = 1$, $j_{\ell}$ is a spherical Bessel function of order $\ell$, $r(z)$ is the comoving distance, $\left\langle T^X\right\rangle(z)$ is the spatially averaged brightness temperature of the spectral line $X$, and $b^X(z)$ is its bias. In Eq.~\eqref{eq:transfer} we neglect nonlinear contributions to clustering and bias, a valid approximation on sufficiently large scales. 

We consider only two contributions to the covariance of the LIM angular power spectrum: sample variance and instrumental noise. Residual foreground contamination is another source of noise, but as this work focuses on the cross-correlations between different lines, we can safely neglect it. 
The instrumental noise power spectrum in a single dish or an interferometer experiment are given by
\begin{equation}
    \begin{split}
        N^{\text{dish}}_{\ell} &= \frac{T_{\text{sys}}^2 \Omega_{\text{field}}}{\Delta\nu t_{\text{obs}} N_{\text{feeds}} N_{\text{pol}} N_{\text{ant}}}, \\
        N^{\text{interf}}_{\ell} &= \frac{T_{\text{sys}}^2 \Omega_{\text{field}} \Omega_{\text{FOV}}}{\Delta\nu \ t_{\text{obs}} N_{\text{feeds}} N_{\text{pol}} n_s},
    \end{split}
\end{equation}
where $T_{\text{sys}}$ is the system temperature, $\Omega_{\text{field}}$ is the solid angle of the sky probed by the survey, $\Delta \nu$ is the bandwidth corresponding to the redshift bin width, $t_{\text{obs}}$ is the observing time, $N_{\text{ant}}$ is the number of antennas with $N_{\text{feeds}}$ detectors each, that measure $N_{\text{pol}} = 1,2$ polarizations. The field of view of an antenna is given by $\Omega_{\text{FOV}} = c^2/(\nu_{\text{obs}} D_{\text{dish}})^2$, and $n_s$ is the average number density of baselines. For a circular array uniformly covered by antennas, $n_s$ is given by \cite{Bull:2014rha}
\begin{equation}
    n_s = \frac{c^2 N_{\text{ant}}(N_{\text{ant}}-1)}{2\pi \nu_{\text{obs}} (D_{\text{max}}^2 - D_{\text{min}}^2)}.
\end{equation}

Thus, the total observed angular auto-power spectrum is defined as
\begin{equation}
    \tilde{C}^{X,Y}_{\ell} \equiv C^{X,Y}_{\ell} + N^{X,Y}_{\ell} \delta_{X,Y}^K,
\end{equation}
where $N^{X, Y}_{\ell}$ is the noise angular power spectrum corresponding to the correlation of $X$ and $Y$, and $\delta^K$ is the Kronecker delta. We note that, since the instrumental noise terms in different LIM surveys are uncorrelated, it is only added to the auto-correlations. 

\section{The Estimator}\label{sec:estimator}
\subsection{Signal model}\label{subsec:signalmodel}
Two-point correlation functions are often assumed to be symmetric under the exchange of radial position. By relaxing this assumption, the angular cross-correlation between tracers X and Y can be separated into symmetric and antisymmetric parts, defined respectively as
\begin{equation}
\begin{split}
    S^{X,Y}_{\ell}(z_i, z_j) &\equiv \frac{1}{2}\left[ C^{X,Y}_{\ell}(z_i, z_j) + C^{X,Y}_{\ell}(z_j, z_i) \right]\\
    A^{X,Y}_{\ell}(z_i, z_j) &\equiv \frac{1}{2}\left[ C^{X,Y}_{\ell}(z_i, z_j) - C^{X,Y}_{\ell}(z_j, z_i) \right].
\end{split}
\label{eq:SandA}
\end{equation}

A variety of contributions to the antisymmetric component have been studied (see, e.g., Ref.~\cite{Bonvin:2013ogt}). We assume for simplicity that the only contribution to the antisymmetric component is the evolution of the bias and global temperature of the cross-correlated fields.  More precisely, the amplitude of the antisymmetric part is proportional to $T^X_{i}b_i^X T^Y_j b^Y_j - T^X_j b^X_j T^Y_i b^Y_i$, where subscripts denote the corresponding redshift bin. This contribution is different from zero if the evolution of the two spectral lines is uneven over redshift. 

We further propose taking the ratio between the antisymmetric and symmetric parts and study the features of this additional estimator, which we define as
\begin{equation}
R^{X, Y}_{\ell}(z_i, z_j)\equiv \frac{A^{X, Y}_{\ell}(z_i, z_j)}{S^{X, Y}_{\ell}(z_i, z_j)}. 
\label{eq:ratio}
\end{equation}

Our main motivation to consider this ratio is its potential to break the degeneracy between astrophysical and cosmological information. The LIM power spectrum carries information about the astrophysical processes that drive the line emission or absorption. However, at linear order, this dependence will only change the amplitude of the power spectrum through the global brightness temperature $\left\langle T^X \right\rangle$ and the bias $b^X$, which are degenerate with the amplitude of the matter power spectrum. The ratio between the antisymmetric and symmetric parts of the cross-correlation can break this degeneracy, since they are two independent tracers of the same underlying matter density field. Furthermore, while on a realization-by-realization basis the measurement of these cross-correlations will be affected by sample variance, their ratio, in the low-noise limit, will not \cite{Seljak:2008xr}. 

\subsection{Covariance}
The covariance for the angular cross-correlation estimator and for its symmetric and antisymmetric components, defined in Eq.~\eqref{eq:SandA}, can be derived through a standard calculation, which we omit here for brevity. The result for the angular cross-correlation is given by
\begin{equation}
\begin{split}
    \text{Cov} &\left[C^{X,Y}_{\ell, (ij)}, C^{X,Y}_{\ell, (pq)}\right] = \\
    & = \frac{\tilde{C}^{X,X}_{\ell, (ip)}\tilde{C}^{Y,Y}_{\ell, (jq)} + \tilde{C}^{X,Y}_{\ell,(iq)} \tilde{C}^{Y,X}_{\ell, (jp)}}{(2\ell +1)f_{\rm sky}},
\end{split}
\end{equation}
where we use $C_{\ell,(ij)}^{X,Y}\equiv C_\ell^{X,Y}(z_i,z_j)$ to compress the notation, and $f_{\rm sky}$ is the fraction of sky probed.\footnote{Note that the spherical harmonic expansion is defined for all sky. If $f_{\rm sky}<1$, the angular power spectrum has mode-coupling, that we omit in this work for simplicity.}

For the antisymmetric component, the covariance is given by
\begin{equation}
\begin{split}
    \text{Cov} &\left[A^{X,Y}_{\ell, (ij)}, A^{X,Y}_{\ell, (pq)}\right] = \frac{1}{4(2\ell +1)f_{\rm sky}} \times\\ \times \Big[ &\tilde{C}^{X,X}_{\ell, (ip)}\tilde{C}^{Y,Y}_{\ell, (jq)} + \tilde{C}^{X,Y}_{\ell,(iq)} \tilde{C}^{Y,X}_{\ell, (jp)} - \tilde{C}^{X,X}_{\ell, (iq)}\tilde{C}^{Y,Y}_{\ell, (jp)}- \\ - &\tilde{C}^{X,Y}_{\ell, (ip)} \tilde{C}^{Y,X}_{\ell, (jq)} - \tilde{C}^{X,X}_{\ell, (jp)}\tilde{C}^{Y,Y}_{\ell, (iq)} - \tilde{C}^{X,Y}_{\ell, (jq)}\tilde{C}^{Y,X}_{\ell, (ip)}+ \\ + & \tilde{C}^{X,X}_{\ell, (jq)}\tilde{C}^{Y,Y}_{\ell,(ip)} + \tilde{C}^{X,Y}_{\ell,(jp)} \tilde{C}^{Y,X}_{\ell,(iq)}\Big].
    \label{eq:covAl}
\end{split}
\end{equation}
The result for the symmetric part can be derived in a similar manner and yields the same expression but with all terms positive. The covariance for the ratio can be found using standard error propagation and is shown in Appendix~\ref{appendix:ratio}. 

\section{Line Models}\label{sec:linemodels}

The HI field is defined as the brightness temperature contrast between neutral hydrogen and the CMB. During the EoR, the gas has been significantly heated and the spin temperature is much higher than the CMB temperature. We therefore make the standard simplifying assumption that the contribution from spin-temperature fluctuations can be neglected \cite{2011MNRAS.411..955M, Zaldarriaga:2003du}. We further simplify the HI brightness temperature by ignoring redshift-space distorsions \cite{2007ApJ...669..663M}. The HI brightness temperature $\delta T^{\HI}$ at a position ${\bf x}$ can be written as
\begin{equation}
\delta T^{\HI} ({\bf x}) = T_0(z) x_{\HI}({\bf x})\left[ 1 + \delta_{\rho}({\bf x}) \right],
\end{equation}
where $T_0 = 27\ $mK$ \left( \frac{\Omega_b h^2}{0.022} \right) \left( \frac{0.14}{\Omega_m h^2}\frac{1+z}{10} \right)^{1/2}$, $x_{\text{HI}}$ is the neutral hydrogen fraction at a position ${\bf x}$, and $\delta_{\rho}$ is the gas density perturbation. Recalling the transfer function defined in Eq.~\eqref{eq:transfer} in the linear regime, we wish to calculate $\left\langle T^{\HI}\right\rangle(z) = T_0(z)\left\langle x_{\HI}\right\rangle(z)$ and $b^{\HI}(z)$.

We adopt a simple model for the average neutral hydrogen fraction as a function of redshift, which is given by~\cite{Pritchard:2010pa,Kovetz:2018zan}
\begin{equation}
\left\langle x_{\text{HI}}\right\rangle (z) = \frac{1}{2} \left[ 1 + \tanh \left( \frac{z -z_r}{\Delta z_r} \right) \right].
\end{equation}
The main features are described by the parameters $z_r$ and $\Delta z_r$, which correspond to the midpoint of reionization and its duration, respectively. We adopt the fiducial values of $z_r=8$ and $\Delta z_r = 1$.

Before the beginning of reionization, the spatial distribution of the neutral hydrogen gas is expected to follow the matter distribution. Reionization begins after the first ionizing sources are formed in high-density regions, giving rise to an anti-correlation between the neutral hydrogen and the matter distribution. This is equivalent to a bias $b^{\text{HI}} \sim 1$ when $\left\langle x_{\text{HI}}\right\rangle \sim 1$ and negative as reionization progresses. We model this behavior using the following parametrization for the linear HI bias
\begin{equation}
b^{\HI}(z) = \eta(\left\langle x_{\text{HI}}\right\rangle (z) -1) +1,
\label{eq:HIbias}
\end{equation}
where a fit to semi-numerical simulations yields $\eta=14.8$ \cite{Hoffmann:2018clb}.

In order to model the CO emission, we assume that the spectral lines are sourced within dark matter halos and that there is a known relation between the mass $M$ of a halo at redshift $z$ and the luminosity $L^{\CO}(M,z)$ of the line emission. Given a halo mass function $dn/dM$, we can compute the expected luminosity density as
\begin{equation}
    \left\langle \rho^{\CO}_L\right\rangle (z) = \int dM \frac{dn}{dM}(M,z) L^{\CO}(M,z).
\end{equation}

To illustrate how our results depend on the highly uncertain astrophysical modelling, we consider two different cases. In one case, we consider a power law for the halo mass-luminosity relation (see e.g., Ref. \cite{Breysse:2015saa}), given by
\begin{equation}
    L^{\CO}(M) = A \left(\frac{M}{M_{\odot}}\right)^b L_{\odot},
    \label{eq:MassPow}
\end{equation}
where we adopt the fiducial values $A=2.8\times 10^{-5}$, and $b=1$.

We also consider the model presented in Ref.~\cite{Li:2015gqa}, hereafter referred to as L16. The approach adopted in L16 is to parametrize the relation between the star-formation rate and the halo mass at a given redshift. The CO luminosity is then empirically inferred from the star-formation rate, with the total infrared luminosity as an intermediate tracer. We highlight that the model is calibrated on empirical correlations observed at much lower redshifts than the ones considered here, and therefore introduce large modelling uncertainties. We use the set of fiducial parameters described in L16 and, for both models, use the {\tt lim} \footnote{https://github.com/pcbreysse/lim} package for the calculations. 

Assuming dark matter halos trace the underlying matter distribution with a linear bias $b(M,z)$,  the bias of the brightness temperature perturbations is then given by the luminosity-averaged bias
\begin{equation}
    b^{\CO}(z)=\frac{\int dM L^{\CO}(M,z) b(M,z)\frac{dn}{dM}(M,z) }{\int dM L^{\CO}(M,z) \frac{dn}{dM}(M,z) }.
\end{equation}

The CO line average brightness temperature at redshift $z$ can  be written in terms of the luminosity density as
\begin{equation}
    \left\langle T^{\CO}\right\rangle (z) = \frac{c^3 (1+z)^2}{8\pi k_B \nu^3 H(z)} \left\langle \rho^{\CO}_L\right\rangle(z),
\end{equation}
where $c$ is the speed of light, $k_B$ is the Boltzmann constant, and $H(z)$ is the expansion rate \cite{Lidz:2011dx}.



\section{Signal-to-noise ratio and forecasts}\label{sec:Forecast}
\begin{figure}[ht]
    \centering
    \includegraphics[width=0.4\textwidth]{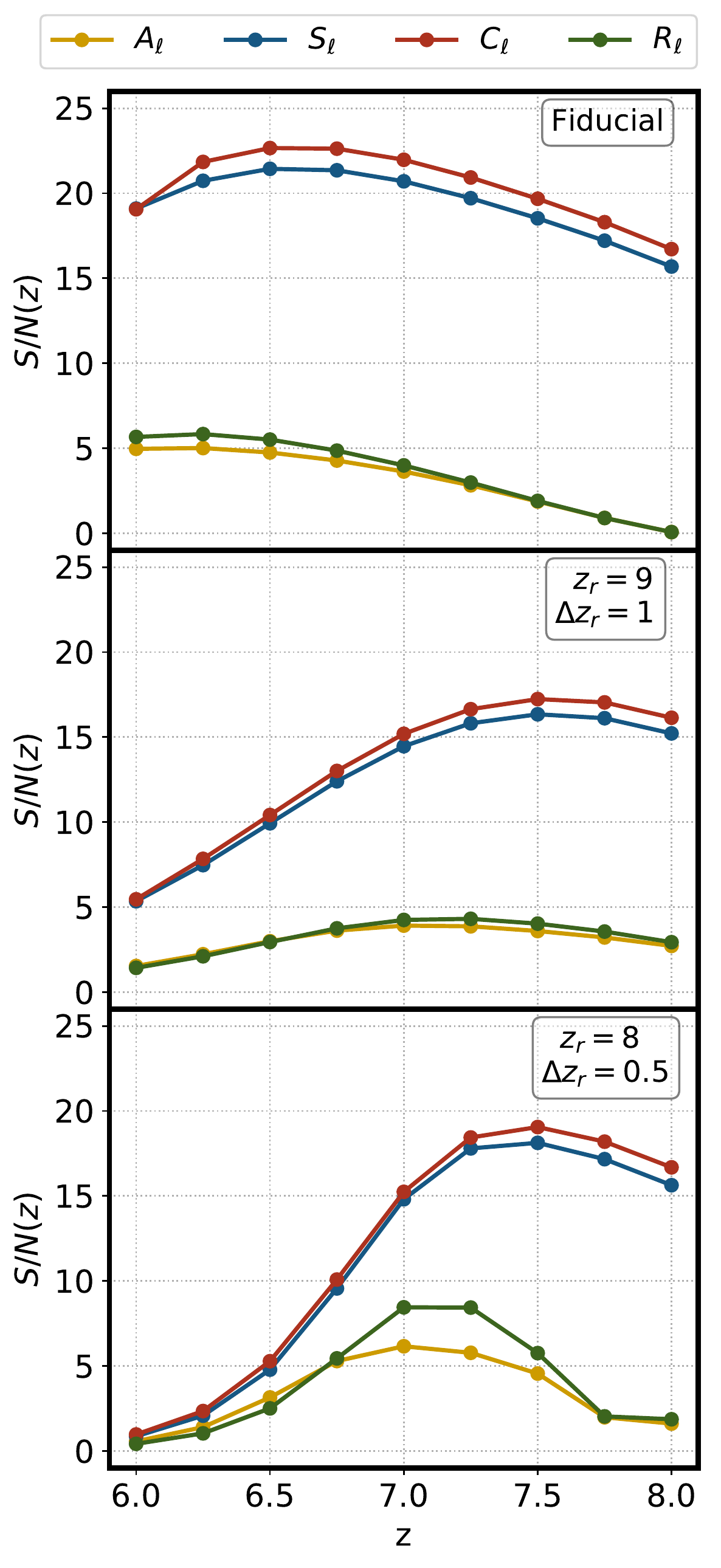}
    \caption{Signal-to-noise ratio as a function of redshift. Each curve corresponds to one of the four different estimators of the cross-correlation between CO and HI. The upper panel shows the signal-to-noise ratio for our fiducial reionization parameters, with $z_r = 8$ and $\Delta z_r = 1$. The middle panel corresponds to a scenario with an earlier reionization, at $z_r = 9$, and the bottom panel corresponds to a faster reionization scenario, with $\Delta z_r = 0.5$.}
    \label{fig:SNR}
\end{figure}

\begin{figure*}
  \centering 
    \includegraphics[width=0.49\textwidth]{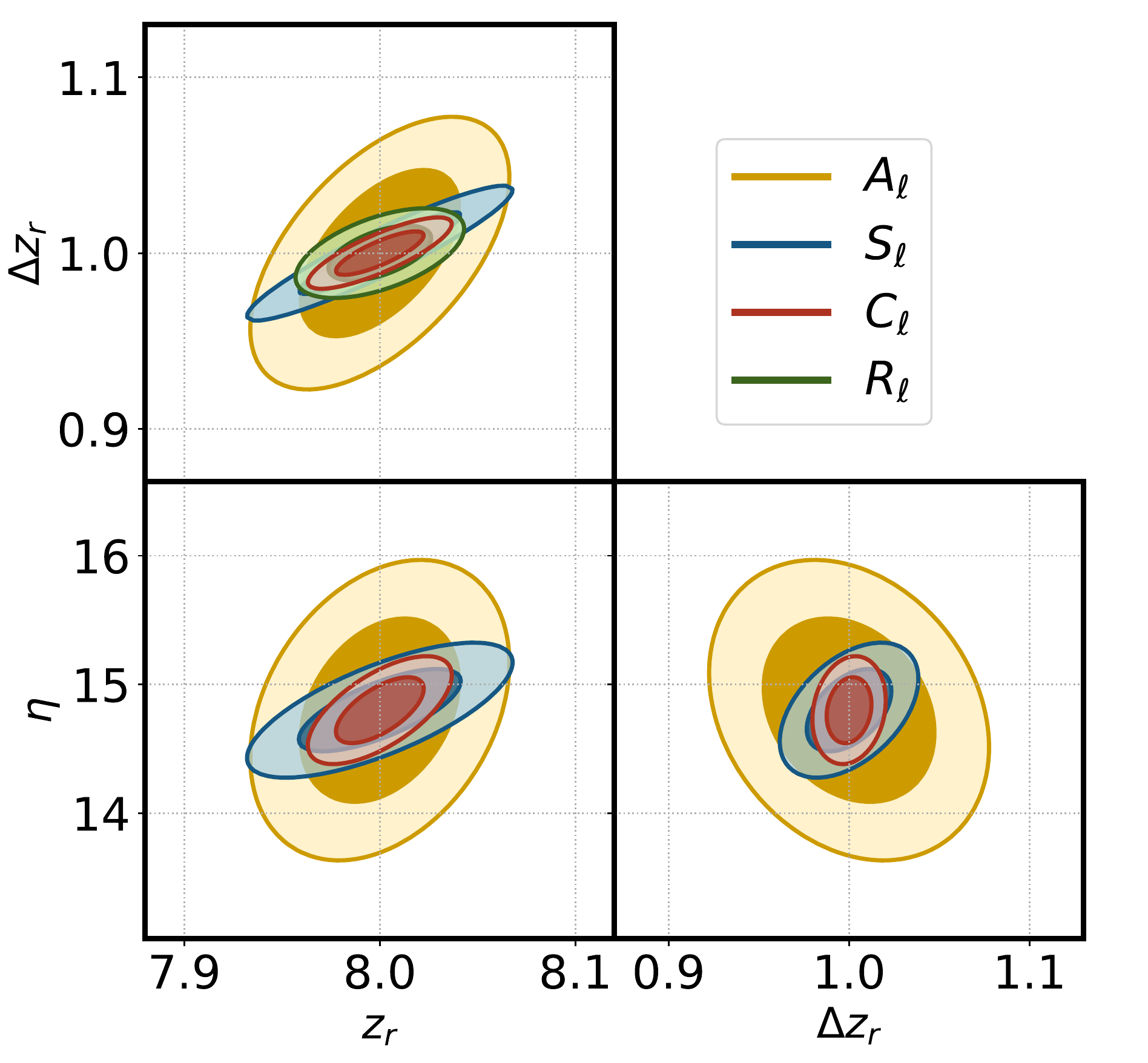}
    \includegraphics[width=0.49\textwidth]{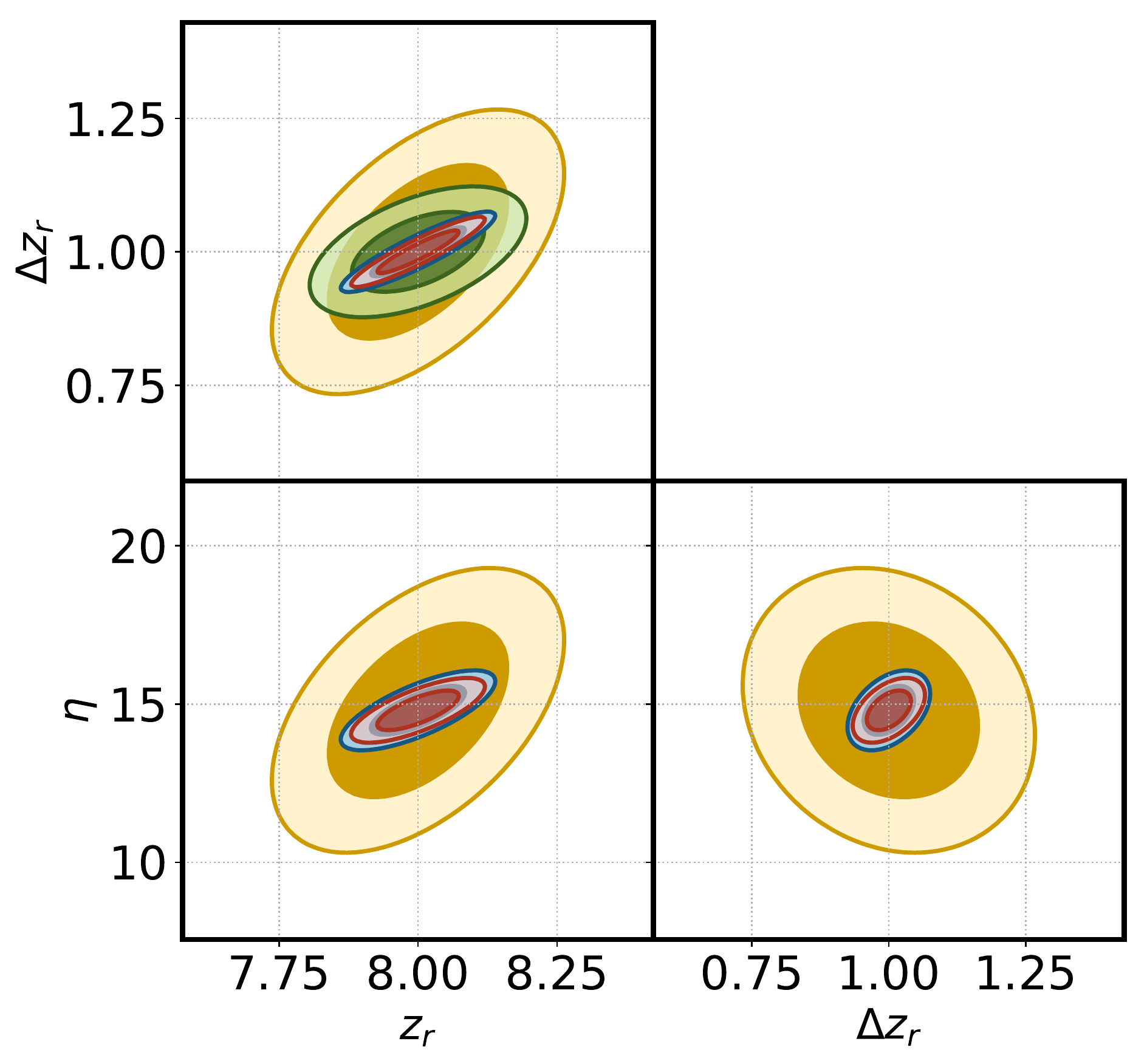}
  \caption{Forecasted marginalized uncertainties on reionization parameters at 68\% and 95\% confidence levels. The panel on the left shows the forecast assuming a power law model for the CO luminosity with fiducial parameters, and the panel on the right corresponds to the model in L16. Each color corresponds to a different estimator of the cross-correlation between CO and HI, defined in Section~\ref{subsec:signalmodel}.}
  \label{fig:fisher}
\end{figure*}

We consider CO and HI surveys that overlap between redshifts $z=6-8$, on a patch of the sky with $\Omega_{\text{field}} = 100$ deg$^2$ and both with 3000 hours of total observing time. We assume that the CO experiment is an array of single dish antennas with a total $t_{\text{obs}} N_{\text{feeds}} N_{\text{pol}} N_{\text{ant}}/T^2_{\text{sys}} = 940$ h/K$^2$ and an angular resolution of $\theta_{\text{FWHM}} = 4$ arcmin. We conceive it as an upgrade of COMAP~\cite{Cleary_COMAP} for the next generation of LIM experiments. 

For the HI survey, we consider an interferometer based on the experimental configuration of HERA \cite{DeBoer:2016tnn}. We assume the array has 350 antennas, with one beam each, dual-polarization, and minimum and maximum baselines of $D_{\text{min}} = 14.6$ m and $D_{\text{max}} = 876$ m. The system temperature is given by $T_{\text{sys}} = 100 + 120(\nu_{\text{obs}}/150\ \text{MHz})^{-2.55}\ $K and $\Omega_{\text{field}}$ is limited to the overlap area with the CO experiment.

We take Gaussian redshift bins of width $\sigma_z = 0.125$, separated by $\Delta z = 0.25$, and include cross-correlations between redshift pairs separated by $\Delta z_{\text{max}}$ up to $0.5$, that is, up to two adjacent bins. For larger radial distances, the correlation due to density clustering is negligible. Due to the limited survey volume, scales larger than $\ell_{\rm min} = \pi/\sqrt{\Omega_{\text{field}}} = 18$ are excluded.

We compute the signal-to-noise ratio $S/N$ for the four estimators considered in this work: the antisymmetric part $A^{\CO, \HI}_{\ell}$ of the cross-correlation, the symmetric part $S^{\CO, \HI}_{\ell}$, the full cross-correlation $C^{\CO, \HI}_{\ell}$ including both symmetric and antisymmetric parts, and the ratio $R^{\CO, \HI}_{\ell}$. We calculate $S/N$ as function of redshift $z_i$ summing over all the redshift bins that cross-correlate with it. For each estimator $E$, this corresponds to
\begin{equation}
    \left(S/N\right)(z_i) = \left[ \sum_{j, \ell}\left(\frac{E^{\CO, \HI}_{\ell, (ij)}}{\sigma_{E_{\ell, (ij)}}}\right)^2 \right]^{1/2},
    \label{eq:snr}
\end{equation}
where $E = A, S, C, R$, and $\sigma^2_{E_{\ell, (ij)}}$ is the variance of the estimator $E$.

The $S/N$ obtained for each estimator using Eq.~\eqref{eq:snr} is shown in Fig.~\ref{fig:SNR}, where the power law model for the CO line was adopted. Each panel in Fig.~\ref{fig:SNR} corresponds to different choices of   reionization parameters. With the fiducial values and a power-law model for CO, both the antisymmetric and ratio estimators present $S/N\gsim 1$ for redshifts below $z\sim 7.5$. The $S/N$ for the symmetric and full estimators are above $1$ across all redshifts. 
 However, this is model dependent: the two lower  panels show how the significance of the signal shifts towards higher redshifts for earlier and faster reionization scenarios, respectively. With the L16 model, the resulting $S/N$ curves have roughly the same redshift dependence, but are $\sim 5$ times lower.


To study the potential advantages of considering the antisymmetric and ratio estimators, we forecast the precision to which each of the four estimators discussed above can constrain the reionization parameters. To do so, we compute the Fisher matrix~\cite{Tegmark_fisher97,Fisher:1935} for the parameters $\theta_{\alpha} ,\theta_{\beta} = z_r, \Delta z_r, \eta $, which is given by
\begin{equation}
\begin{split}
F_{\alpha \beta} & =  \sum_{i,j,p,q, \ell} \frac{\partial E^{\CO, \HI}_{\ell, (ij)}}{\partial \theta_{\alpha}}\times \\\times&\text{Cov}^{-1}\left[E^{\CO, \HI}_{\ell, (ij)}, E^{\CO, \HI}_{\ell, (pq)}\right]\times \frac{\partial E^{\CO, \HI}_{\ell, (pq)}}{\partial \theta_\beta} .
\end{split}
\end{equation}

We show in Fig.~\ref{fig:fisher} the marginalized constraints on reionization parameters at 68\% and 95\% confidence levels, for both the power law model (left) and the model from L16 (right). Fig.~\ref{fig:fisher} shows that the antisymmetric and symmetric components have a high degree of complementarity due to their different degeneracies. Considering both contributions to the cross-correlation therefore leads to a significant improvement on the constraints. We find that the marginalized constraints on both $z_r$ and $\Delta z_r$ for the full cross-correlation estimator are improved by $45\%$ relative to the symmetric part only, and $20\%$ for $\eta$. It is important to highlight, however, that these values are highly model dependent.

We note that the constraint from the ratio estimator is only shown in the top panel since $\eta$ is very poorly constrained. This can be understood from Eq.~\eqref{eq:HIbias}. Since both the numerator and denominator in the ratio estimator have a factor of $\eta$, at a low neutral hydrogen fraction, $\eta$ approximately cancels out.

\begin{figure*}
    \centering
    \includegraphics[width=\textwidth]{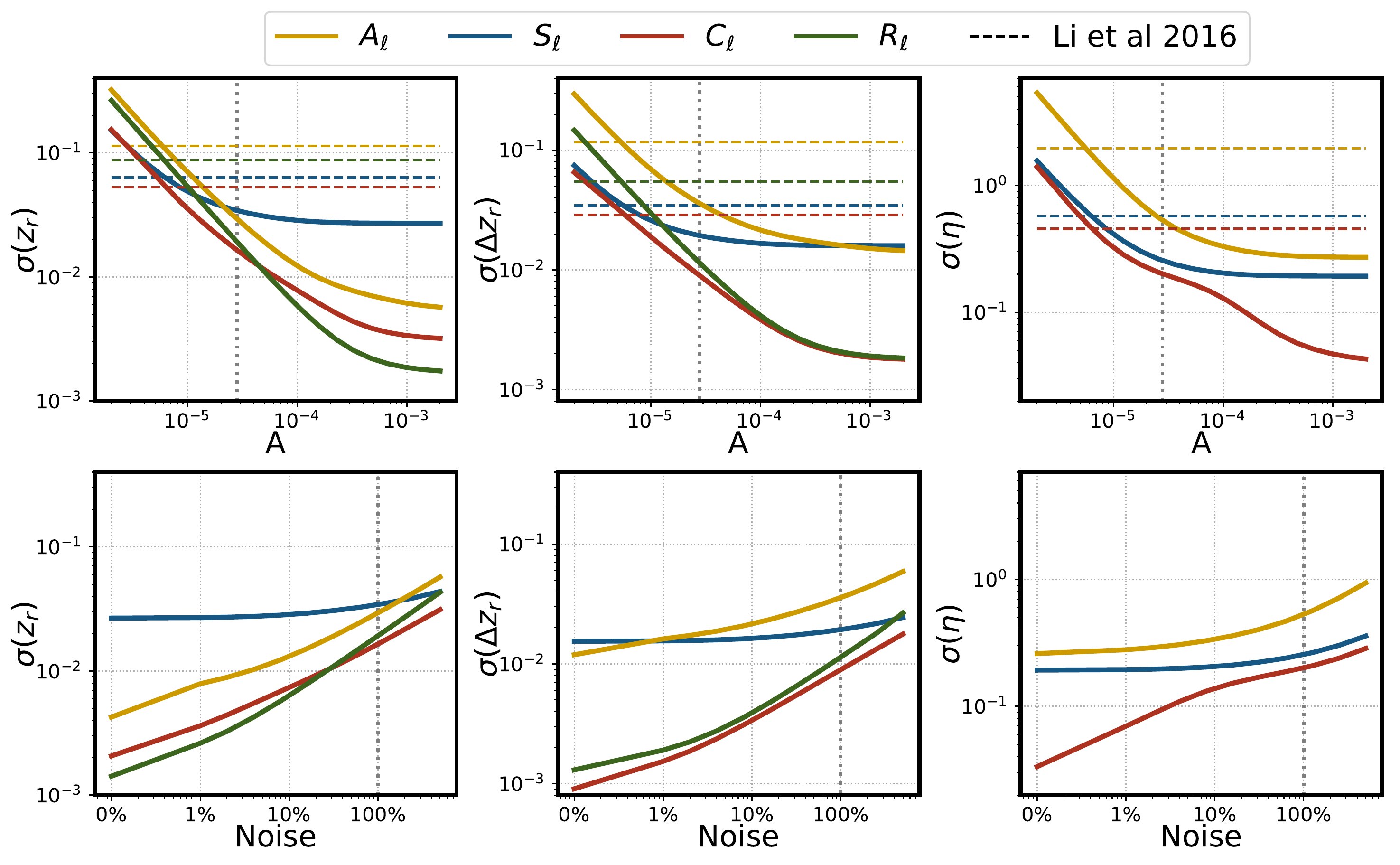}
    \caption{Marginalized 68\% confidence-level forecast as a function of the amplitude of the CO luminosity-halo mass relation (top row) and as function of the instrumental noise relative to the fiducial value(bottom row). In all six panels, the solid curves show the forecast when the power law model for the CO luminosity is adopted, and each color corresponds to a different estimator. On the top row, the dashed horizontal lines show the forecast for the L16 model and the dotted vertical lines mark the fiducial value of A used in this work. Notice that the y-axis scale is different for each column.}
    \label{fig:astro_noise}
\end{figure*}

For the L16 model, Fig.~\ref{fig:fisher} shows that not only are all forecasts less constraining relative to the power law model, but also that the {\it relative} information in the antisymmetric part is significantly reduced. This is mostly due to the lower predicted global brightness temperature relative to the power law model.  Comparing once again the marginalized constraints from the full cross-correlation with the symmetric part only, we find an improvement of $13\%$ on both $z_r$ and $\Delta z_r$, and $19\%$ for $\eta$.

To further investigate how our results depend on the signal and noise amplitudes, we show in Fig.~\ref{fig:astro_noise} the marginalized 68\% confidence-level forecast as a function of the amplitude $A$ of the luminosity-halo mass relation (defined in Eq.~\eqref{eq:MassPow}) and the amount of instrumental noise relative to the fiducial value. The first row in Fig.~\ref{fig:astro_noise} shows that the amplitude of the cross-correlation signal can change the relative information in the symmetric and antisymmetric parts. For a lower signal, or equivalently, higher noise (bottom row), the symmetric part is shown to carry most of the information in cross-correlation. However, for higher signal or lower noise, the information in the antisymmetric part increases and can become dominant.

In summary, the gain in considering both symmetric and antisymmetric contributions to the cross-correlation is highly dependent on the astrophysical modelling and instrumental noise, and can be significant in realistic configurations. Particularly for futuristic LIM experiments, with lower instrumental noise, the antisymmetric component of the cross-correlation may carry most of the information about the redshift of reionization and its duration, as shown in the bottom left and middle panels of Fig.~\ref{fig:astro_noise}. The bottom left panel also shows that in the cosmic variance limit, the ratio estimator is more sensitive to the redshift of reionization than the full cross-correlation.

\section{Conclusions}
\label{sec:conclusions}
Line intensity mapping is a promising technique to study cosmology and astrophysics in new regimes. The EoR is a prime target for upcoming intensity mapping experiments, which are expected to elucidate key features of this period of the Universe. With a variety of proposed experiments targeting different atomic and molecular lines, we focus on the potential cross-correlation between different tracers. This has been studied as a complementary probe of reionization and as a way to mitigate foreground contamination and other systematic effects.

In this work we proposed a new estimator for the angular power spectra between two different spectral lines: the antisymmetric cross-correlation. A significant signal is expected during reionization, since this estimator is sensitive to the difference in the redshift evolution of the temperature and bias of the cross-correlated spectral lines. Furthermore, the antisymmetric cross-correlation is likely to be less subject to potential foreground residuals or observational systematics, since these would mostly contribute to the symmetric component. We also studied the ratio between the antisymmetric and symmetric components, motivated by its potential to break the degeneracy between astrophysics and cosmology. 

We focused on the cross-correlation between CO and HI to probe the EoR, but emphasize that the same technique could be applied to any two lines, as long as they evolve unevenly over redshift. A similar analysis could also be applied to lower redshifts, for example at $z\sim$ 2--3 to probe the star formation rate and its dependence with redshift, which we leave for future work.

We studied the detectability of the antisymmetric and ratio estimators for different reionization histories. Assuming a power-law model for the CO luminosity, with the fiducial parameters defined in Section~\ref{sec:linemodels}, we predicted the signal-to-noise ratio for a given instrumental configuration. For the next-generation CO experiment described in Section~\ref{sec:Forecast}, we found $S/N>1$ for all redshifts below $z \lesssim 7.5$.

We estimated the precision to which the antisymmetric cross-correlation and the antisymmetric-to-symmetric ratio can constrain the parameters that specify the reionization history. In order to quantify the gain offered by the proposed estimators, we compared them to the constraints from the symmetric component and the full cross-correlation, which includes both symmetric and antisymmetric parts. We showcase this comparison for two standard models for the CO luminosity. For the two cases considered, we find that the constraints on reionization parameters are improved by 20--45\% and 13--19\% in the full cross-correlation relative to the symmetric-only. 

While the precise gain from the antisymmetric cross-correlation depends on the highly uncertain astrophysical modeling, we have shown that it can be significant in most standard scenarios and that the antisymmetric cross-correlation can be an important complementary probe. We found, in particular, that for futuristic LIM surveys with lower instrumental noise, the antisymmetric cross-correlation provides stronger constraints for the central redshift and duration of reionization than the symmetric part. Furthermore, we showed that the ratio estimator is more sensitive to the central redshift of reionization than the full cross-correlation in the cosmic variance limit.

We envision that the estimators proposed in this work will be of great use to fully accomplish the potential of forthcoming LIM experiments and maximize the information gain about the EoR and the star formation during the epoch of galaxy assembly.

\acknowledgments
 We acknowledge Yi Mao for suggesting the 21-cm--CO antisymmetric cross-correlation as an EoR probe in his talk at the 2019 LIM conference held at the CCA, NYC.
 We thank Patrick C.~Breysse and Dongwoo T.~Chung for useful discussions. This work was supported at Johns Hopkins by NASA Grant No.\ NNX17AK38G, NSF Grant No.\ 1818899, and the Simons Foundation. JLB is supported by the Allan C. and Dorothy H. Davis Fellowship. EDK is supported by a Faculty Fellowship from the Azrieli Foundation.

\bibliography{ref.bib}
\bibliographystyle{utcaps}

\appendix
\section{Ratio Covariance}\label{appendix:ratio}
Using standard error propagation, we compute the covariance for the ratio estimator defined in Eq.~\eqref{eq:ratio}, which is given by
\begin{equation}
\begin{split}
    \text{Cov} &\left[R^{X,Y}_{\ell, (ij)}, R^{X,Y}_{\ell, (pq)}\right] = \frac{\text{Cov} \left[A^{X,Y}_{\ell, (ij)}, A^{X,Y}_{\ell, (pq)}\right]}{S^{X,Y}_{\ell,(ij)} S^{X,Y}_{\ell,(pq)}} + \\ & + \frac{A^{X,Y}_{\ell,(ij)} A^{X,Y}_{\ell,(pq)}}{\left(S^{X,Y}_{\ell,(ij)} S^{X,Y}_{\ell,(pq)}\right)^2} \text{Cov} \left[S^{X,Y}_{\ell, (ij)}, S^{X,Y}_{\ell, (pq)}\right] - \\ & - \frac{A^{X,Y}_{\ell,(ij)}} {\left(S^{X,Y}_{\ell,(ij)}\right)^2 S^{X,Y}_{\ell,(pq)}} \text{Cov} \left[S^{X,Y}_{\ell, (ij)}, A^{X,Y}_{\ell, (pq)}\right] - \\ & - \frac{A^{X,Y}_{\ell,(pq)}} {S^{X,Y}_{\ell,(ij)} \left(S^{X,Y}_{\ell,(pq)}\right)^2} \text{Cov} \left[A^{X,Y}_{\ell, (ij)}, S^{X,Y}_{\ell, (pq)}\right],
\end{split}
\end{equation}
where the covariances between $A^{X,Y}_{\ell, (ij)}$ and $S^{X,Y}_{\ell,(pq)}$, and $S^{X,Y}_{\ell, (ij)}$ and $A^{X,Y}_{\ell,(pq)}$ are given by
\begin{equation}
\begin{split}
    \text{Cov} &\left[A^{X,Y}_{\ell, (ij)}, S^{X,Y}_{\ell, (pq)}\right] = \frac{1}{4(2\ell +1)f_{\rm sky}} \times\\ \times \Big[ &\tilde{C}^{X,X}_{\ell, (ip)}\tilde{C}^{Y,Y}_{\ell, (jq)} + \tilde{C}^{X,Y}_{\ell,(iq)} \tilde{C}^{Y,X}_{\ell, (jp)} + \tilde{C}^{X,X}_{\ell, (iq)}\tilde{C}^{Y,Y}_{\ell, (jp)}+ \\ + &\tilde{C}^{X,Y}_{\ell, (ip)} \tilde{C}^{Y,X}_{\ell, (jq)} - \tilde{C}^{X,X}_{\ell, (jp)}\tilde{C}^{Y,Y}_{\ell, (iq)} - \tilde{C}^{X,Y}_{\ell, (jq)}\tilde{C}^{Y,X}_{\ell, (ip)}- \\ - & \tilde{C}^{X,X}_{\ell, (jq)}\tilde{C}^{Y,Y}_{\ell,(ip)} - \tilde{C}^{X,Y}_{\ell,(jp)} \tilde{C}^{Y,X}_{\ell,(iq)}\Big].
\end{split}
\end{equation}
and
\begin{equation}
\begin{split}
    \text{Cov} &\left[S^{X,Y}_{\ell, (ij)}, A^{X,Y}_{\ell, (pq)}\right] = \frac{1}{4(2\ell +1)f_{\rm sky}} \times\\ \times \Big[ &\tilde{C}^{X,X}_{\ell, (ip)}\tilde{C}^{Y,Y}_{\ell, (jq)} + \tilde{C}^{X,Y}_{\ell,(iq)} \tilde{C}^{Y,X}_{\ell, (jp)} - \tilde{C}^{X,X}_{\ell, (iq)}\tilde{C}^{Y,Y}_{\ell, (jp)}- \\ - &\tilde{C}^{X,Y}_{\ell, (ip)} \tilde{C}^{Y,X}_{\ell, (jq)} + \tilde{C}^{X,X}_{\ell, (jp)}\tilde{C}^{Y,Y}_{\ell, (iq)} + \tilde{C}^{X,Y}_{\ell, (jq)}\tilde{C}^{Y,X}_{\ell, (ip)}- \\ - & \tilde{C}^{X,X}_{\ell, (jq)}\tilde{C}^{Y,Y}_{\ell,(ip)} - \tilde{C}^{X,Y}_{\ell,(jp)} \tilde{C}^{Y,X}_{\ell,(iq)}\Big].
\end{split}
\end{equation}
The covariance for $A^{X,Y}_{\ell}$ is given in Eq.~\ref{eq:covAl} and the expression for $S^{X,Y}_{\ell}$ is the same, but with all terms positive.

\bigskip

\end{document}